\documentclass[runningheads]{llncs}
\usepackage[T1]{fontenc}
\usepackage[misc]{ifsym}
\usepackage{graphicx}
\usepackage[pagebackref=true,breaklinks=true,colorlinks,bookmarks=false]{hyperref}
\begin{document}
\titlerunning{An Ensemble of 2.5D ResUnet Based Models}
\title{An Ensemble of 2.5D ResUnet Based Models for Segmentation of Kidney and Masses}
%
%
\author{Cancan Chen$^{1}$ \and Rongguo Zhang$^{1,2}$}
%
%
\authorrunning{C. Chen, et al}

\institute{$^{1}$Infervision Advanced Research Institute, Beijing, China \\
$^{2}$Academy for Multidisciplinary Studies, Capital Normal University, Beijing, China \\
\email{chencancan1018@163.com, zrongguo@cnu.edu.cn}}
\maketitle              
\begin{abstract}
The automatic segmentation of kidney, kidney tumor and kidney cyst on Computed Tomography (CT) scans is a challenging task due to the indistinct lesion boundaries and fuzzy texture. Considering the large range and unbalanced distribution of CT scans' thickness, 2.5D ResUnet are adopted to build an efficient coarse-to-fine semantic segmentation framework in this work. A set of 489 CT scans are used for training and validation, and an independent never-before-used CT scans for testing. Finally, we demonstrate the effectiveness of our proposed method. The dice values on test set are 0.954, 0.792, 0.691, the surface dice values are 0.897, 0.591, 0.541 for kidney, tumor and cyst, respectively. The average inference time of each CT scan is 20.65s and the max GPU memory is 3525MB. The results suggest that a better trade-off between model performance and efficiency. 


\end{abstract}

\section{Introduction}
In recent years, over 430,000 people are diagnosed with kidney cancer and roughly 180,000 deaths are caused by kidney cancer annually \cite{Sung2021improved}. Kidney tumors are found in an even larger number each year, and in most circumstances, it's not currently possible to radiographically determine whether a given tumor is malignant or benign \cite{De2017improved}. Computer tomography (CT) scans is an import clinical tool to diagnose and detect kidney tumors. Surgery is the most common treatment option. Radiologists and surgeons are also dedicated to study kidney tumors on CT scans to design optimal treatment schedule by annotating the kidney and its masses manually. However, the manual annotation is a repetitive heavy laborious work and always subjective and varied from the different radiologists. Considering this, automatic segmentation of kidney and kidney tumors is a promising tool for alleviating these clinical problems.

Based on the 2019 and 2021 Kidney Tumor Segmentation Challenge \cite{KiTS,KiTSDataset}, KiTS23 features an expanded training set (489 cases) with a fresh never-before-used test set (110 cases), and aims to serve a stronger benchmark and develop the best automatic semantic segmentation system for kidney tumors. Besides, hardware (GPU, CPU, etc)about average inference time of each case are also real factors in clinical application scenes, so it is important to balance the performance and efficiency of the automatic semantic segmentation system.

In this paper, based on the original ResUnet \cite{diakogiannis2020resunet}, we propose an efficient coarse-to-fine semantic segmentation framework to automatically segment kidneys and tumors. In the coarse segmentation stage, the whole CT images are re-sampled to $128\times128\times128$ as the input. In the fine segmentation stage, we firstly obtain regions of interest (ROIs) for the kidney on the whole CT images based on the coarse segmentation mask, and according to this, randomly crop cubes along z-axis, which are re-sampled to $48\times224\times384$ as the input. Besides, a cascaded model, consisting of the kidney segmentation model and the kidney-tumor-cyst segmentation model, is applied on the second stage.

The main contributions of this work are summarized as follows:
\begin{itemize} 
\item We propose a coarse-to-fine semantic segmentation framework, which can effectively segment kidney, kidney tumor and kidney cyst from the abdominal CT images.

\item We firstly conduct a statistical analysis on the spacing resolution of all CT images, especially the thickness distribution at the z-axis, which sparks the major design ideas about the random cropping method, patch size and 2.5D ResUnet structure on the fine segmentation stage.

\item We evaluate our proposed framework by 5-fold cross validation on Kits23 data set.
\end{itemize}

\section{Methods}

Semantic segmentation of organs and lesions is a common task for medical image analysis. There are already numerous accurate and efficient algorithms for medical image segmentation, such as U-Net \cite{ronneberger2015u}, ResUNet \cite{diakogiannis2020resunet}, nnU-Net \cite{isensee2021nnu}, et al. Based on the natural properties of Kits23 CT images and the strong baseline\cite{KiTS}, we develop a whole-volume-based coarse-to-fine framework as follows, which consists of coarse segmentation, fine kidney segmentation (two-classification task of kidney and others) and fine tumor-mass segmentation (three-classification task of tumor, cyst and other kidney regions).

\begin{figure}[!htbp]
\centering
\includegraphics[width=1.0\textwidth]{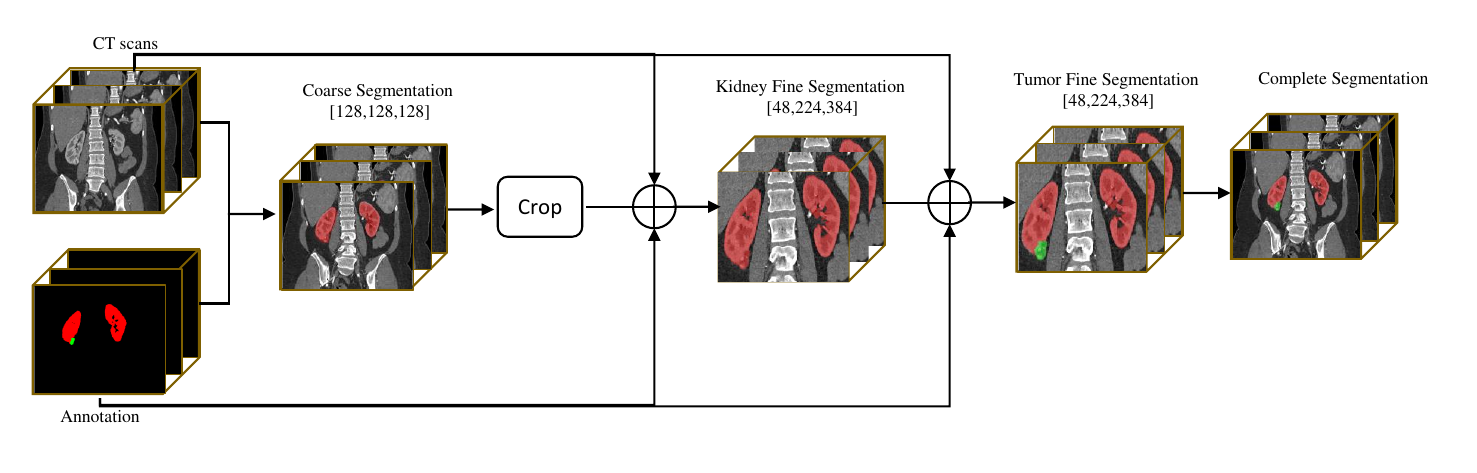}
\caption{An overview of our coarse-to-fine segmentation framework.}
\label{fig:framework}
\end{figure}

\subsection{Preprocessing}
Our proposed method includes the following preprocessing steps:
\begin{itemize} 
\item Cropping strategy: 

In the coarse segmentation stage, the input is the whole volumes. In the fine segmentation stage, the kidney ROIs are firstly cropped from the whole volumes based on the coarse segmentation mask, and after that, we randomly crop 3D cubes from the kidney ROIs only along z-axis to ensure 2D kidney scans' structural integrity.

\item Re-sampling method for anisotropic data: 
 
The original images are re-sampled to $128 \times 128 \times 128$ for coarse segmentation. In the fine segmentation stage, if the shape of the cropped kidney ROI is $d \times w \times h$, it will be resampled to $d \times 224 \times 384$(in this work, $d=48$), i.e., no-re-sampling at z-axis direction, and re-sampling at x/y-axis direction due to the shape distribution of all kidneys.

\item Intensity normalization method:
 
Images are clipped to range [-200, 400] and normalized to range [-1, 1].
 
 \item Others: 
 
To improve the training and testing efficiency, mixed precision is adopted in the whole process of our framework working. 
\end{itemize}

\subsection{Proposed Method}
Our proposed framework is shown in Figure~\ref{fig:framework}. The details of two stages are addressed as follows.
\subsubsection{Coarse Segmentation}
We firstly use a original ResUnet~\cite{diakogiannis2020resunet} to obtain the coarse segmentation mask of all kidneys, and the input size is $128\times128\times128$. The kidney tumor and masses are always located in the kidney region. Based on this, the kidney ROI of each CT image is cropped as the input of the next segmentation stage. This step reduces the computational cost of irrelevant information on this task and preserves all segmentation target. 

\subsubsection{Fine Segmentation}
\begin{figure}[!htbp]
\centering
\includegraphics[width=\textwidth]{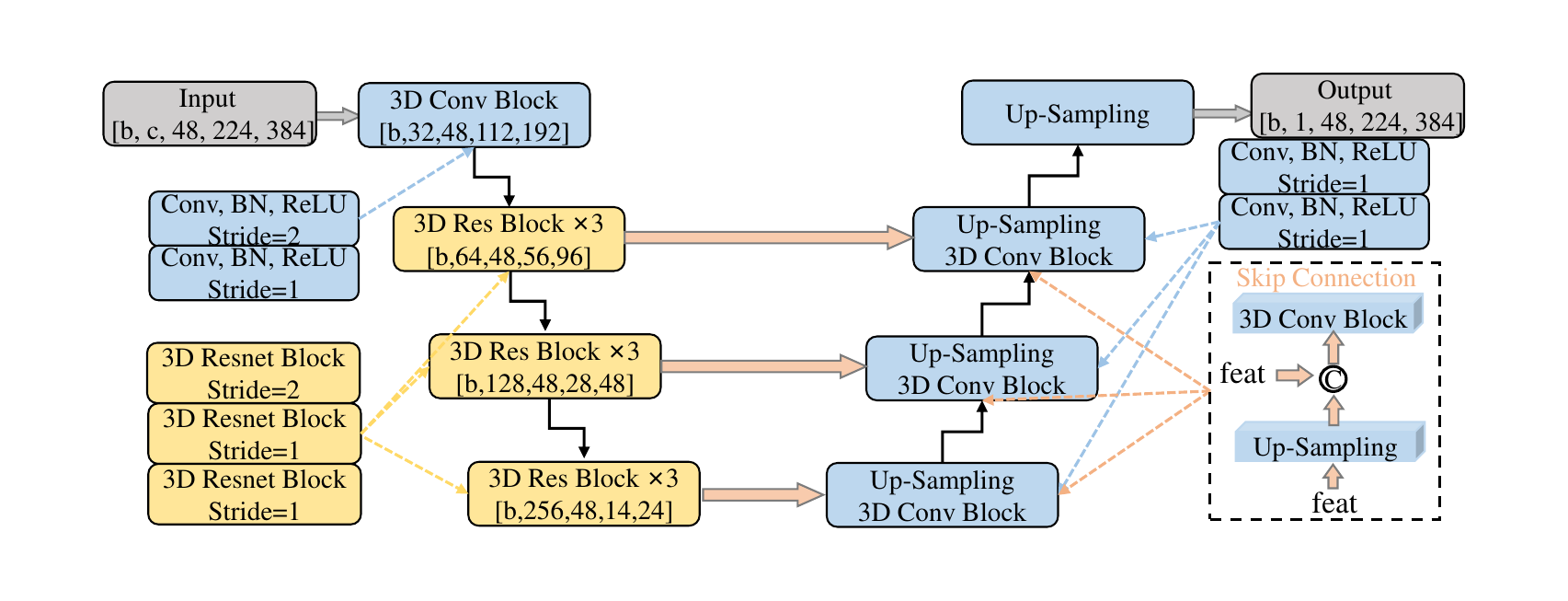}
\caption{Our proposed network architecture.}
\label{fig:network}
\end{figure}

The fine segmentation consists of kidney fine segmentation and lesion fine segmentation. Notably, the thickness range of all CT scans is between 0.5mm and 5mm. To resolve the data heterogeneity, cropping or re-sampling should be used. Considering the framework efficiency, the kidney ROIs are re-sampled to the fixed size at x and y direction, and then, we crop the cubes from kidney ROIs only along z axis. That's to say, if the shape of the kidney ROI is $d \times w \times h $, it will be re-sampled to $d \times 224 \times 384$, and the re-cropped cube size for fine segmentation is $48 \times 223 \times 384$ in this work. Finally, we adopt 2.5D ResUnet as the segmentation backbone. Network architecture has 3 down-sample layers, 3 up-sample layers, and no down-sample at z direction for the high-performance and high-efficiency of our framework, which is shown in \ref{fig:network}.

\subsubsection{Loss function}
We use the summation of the weighted Dice loss and Cross-Entropy loss as the final compound loss function which has been proved to be robust in various medical image segmentation tasks~\cite{LossOdyssey}. 

\subsubsection{Other tricks} 
The mixup \cite{zhang2017mixup} and hard examples mining are adopted in the model training process, both of which significantly improve the ResUnet's fitting capability. 

\subsection{Post-processing}
In the inference process, the connected component analysis~\cite{cc3d2021} is applied to avoid the influence of noise. Based on the natural attributes of the kidney and lesions, we choose connected component regions larger than 10000 pixels as the final segmentation results. Notably, we abandon the multi-models ensemble method for efficient inference. Our method consists of the coarse segmentation model, the kidney fine segmentation model (background, kidney) and the lesion fine segmentation model (kidney, cyst, tumor). The final result is the average of the two predictions for the original image and the mirror image along the z-axis.

\section{Results}
\subsection{Dataset and evaluation measures}
The KiTS23 organizer has publicly released an expanded training set, totally 489 cases, based on  KiTs19 and KiTs21. The volumetric Dice coefficient and the Surface Dice are used to evaluate algorithms, and the following Hierarchical Evaluation Classes (HECs) will be used: Kidney + Tumor + Cyst, Tumor + Cyst and Tumor only.

\subsection{Implementation details}
\subsubsection{Environment settings}
The development environments and requirements are presented in Table~\ref{table:env}.

\begin{table}[!htbp]
\caption{Development environments and requirements.}\label{table:env}
\centering
\begin{tabular}{ll}
\hline
Windows/Ubuntu version       & Ubuntu 18.04.06 LTS\\
\hline
CPU   & Intel(R) Core(TM) i9-10900X CPU @ 3.70GHz \\
\hline
RAM                         &96GB\\
\hline
GPU (number and type)                         & Four NVIDIA RTX A4000 16G\\
\hline
CUDA version                  & 11.5\\                          \hline
Programming language                 & Python 3.7\\ 
\hline
Deep learning framework & Pytorch (Torch 1.7.1+cu110, torchvision 0.8.2) \\
\hline
Specific dependencies         &                        \\                                                                      
\hline
(Optional) Link to code     &                                                                \\
\hline
\end{tabular}
\end{table}

\subsubsection{Training protocols}
In our training process, we performed the following data augmentation with project MONAI~\cite{monai2022} : 1). randomly crop the volumes with range $[0.6, 1.3]$; 2). add brightness, contrast and gamma augmentation on the volumes and lesions with range $[0.6, 1.5]$, respectively. 3). random elastic transform with prob=0.5 and with sigma from range 3 to 5 and magnitude from range 100 to 200; 4). clip volumes to range $[-1, 1]$. Details of our training protocols are shown in Table~\ref{table:training} and Table~\ref{table:training2nd}.

\begin{table*}[!htbp]
\caption{Training protocols for coarse segmentation.}
\label{table:training}
\begin{center}
\begin{tabular}{ll} 
\hline
Network initialization         & ``he" normal initialization\\
\hline
Batch size                    & 4 \\
\hline 
Patch size & 128$\times$128$\times$128  \\ 
\hline
Total epochs & 300 \\
\hline
Optimizer          & ADAMW~\cite{loshchilov2017decoupled} ($weight decay=1e-4$)          \\ \hline
Initial learning rate (lr)  & 1e-4 \\ \hline
Lr decay schedule & CosineAnnealing \\
\hline
\end{tabular}
\end{center}
\end{table*}

\begin{table*}[!htbp]
\caption{Training protocols for fine segmentation.}
\label{table:training2nd}
\begin{center}
\begin{tabular}{ll} 
\hline
Network initialization         & ``he" normal initialization\\
\hline
Batch size                    & 4 \\
\hline 
Patch size & 48$\times$224$\times$384  \\ 
\hline
Total epochs & 600 \\
\hline
Optimizer          & ADAMW~\cite{loshchilov2017decoupled} ($weight decay=1e-4$)          \\ \hline
Initial learning rate (lr)  & 1e-4 \\ \hline
Lr decay schedule & CosineAnnealing \\
\hline
\end{tabular}
\end{center}
\end{table*}

\subsection{Results on cross validation and test data}
Originally, our proposed framework would be evaluated by 5-fold cross validation. However, we only train and evaluate model on fold-0 data set due to the time and computation resource constraints, and all scores are listed in Table \ref{table:results}. The average inference time on fold-0 validation (98 cases) and test set (110 cases) is 19.22s and 20.65s, respectively. The max GPU memory at inference step is 3525MB.

\begin{table*}[!htbp]
\caption{Results of our proposed method on fold-0 and test set.}
\label{table:results}
\begin{center}
\begin{tabular}{ccccccccccccccccccccc} 
\hline
Targets & Dice(Fold-0) & Surface Dice(Fold-0) & Dice(Test) & Surface Dice(Test)\\
\hline
Kidney   & $0.9661$  &  $0.9408$ & $0.954$  &  $0.897$\\
Cyst  & $0.8580$  &  $0.7365$ & $0.792$  &  $0.591$\\
Tumor  &$0.8591$  &  $0.7329$ & $0.691$  &  $0.541$\\
\hline
\end{tabular}
\end{center}
\end{table*}

\section{Conclusion}
Based on 2.5D ResUnet, we propose a efficient coarse-to-fine framework for the automatic segmentation of kidney and masses. The experimental results indicate that our framework is effective, but the segmentation robustness of kidney tumors and cysts need further improvement. One possible reason is that the capability level of single model has a lower upper-limit for the hard segmentation task. Thus, the ensemble of multi-models is an alternative solution after balancing performance and efficiency. 

%
%
%
\bibliographystyle{splncs04}
\bibliography{ref}

\end{document}